\documentstyle[12pt]{article}

\begin{document}
\title{ Quasi - relativistic harmonic oscillator bound - states}
\author{ Omar Mustafa and Maen Odeh\\
 Department of Physics, Eastern Mediterranean University\\
 G. Magusa, North Cyprus, Mersin 10 - Turkey\\
 email: omustafa.as@mozart.emu.edu.tr\\
\date{}\\}
\maketitle
\newpage
\begin{abstract}
{\small The quasi - relativistic harmonic oscillator bound - states
constructed by Znojil ( 1996 J. Phys. {\bf A29} 2905) are investigated
via a new methodical proposal. Compared to those obtained by an anonymous
referee ( from a direct numerical integration method) of Znojil's paper [3],
our results appeared to be more favorable than those obtained by Znojil via
quasi - perturbative, variational, Hill - determinant and Riccati - Pad\'{e}
methods. Bound - states with larger angular momenta $l$ are also
constructed.}
\end{abstract}
\newpage

Among exactly soluble Hamiltonians exists the harmonic oscillator
(HO) Hamiltonian\\
\begin{equation}
H^{(HO)}=\frac{{\bf{p}}^2}{2m} + \frac{1}{2}m \omega^2 {\bf{r}}^2.
\end{equation}\\
The equidistant  form of its spectrum have attracted attention in
quantum control theory [1] and represents, in view of [2], a ' long
sought - after dream' of ' steering wavepackets into desired states '.
Moreover, it fits, via experimental observations, vibrational
excitations of molecules and some low - lying energy levels in atomic
nuclei.

On the other hand, a free spin - 0 field fulfills Klein - Gordon
equation\\
\begin{equation}
(E^2 - H^2)\Psi=0~~~~;~~~H=\sqrt{c^2{\bf p}^2+m^2c^4}, \nonumber
\end{equation}\\
or the Schr\"odinger formulation of it $(E-H)\Psi=0$. Of course there
exist two admissible solutions for a given momentum ${\bf p}$, i.e.
positive  and/or negative  energy solutions.
However, we devote our work to the positive solution in the Schr\"odinger
- formulated Klein - Gordon equation. With the minimal coupling, a Lorentz
4 - vector harmonic oscillator potential is coupled as the 0 - component of
the four - vector potential, i.e. $E \longrightarrow E-eA_0$ with
$eA_0=m\omega^2r^2/2$. Hence, a fully relativistic description of
$H^{(HO)}$ leads to the emergence of a quasi - relativistic harmonic
oscillator (QHO) Hamiltonian\\
\begin{equation}
H^{(QHO)}=\sqrt{m^2c^4 + {\bf{p}}^2c^2} + \frac{1}{2} m \omega^2 {\bf r}^2.
\end{equation}\\
Which, in momentum representation, implies the one - dimensional
Schr\"odinger equation\\
\begin{equation}
\frac{1}{2} m \hbar^2 \omega^2 \left[ -\frac{d^2}{dp^2}
+\frac{l(l+1)}{p^2} \right] \Psi (p)
+[\sqrt{m^2c^4 + {\bf{p}}^2c^2} - mc^2] \Psi (p) = \acute{E} \Psi (p),
\end{equation}\\
where ${\bf{r}}^2=-\hbar^2 \Delta_{\bf p}$, $p\in(0,\infty)$, $\acute{E}$
is the binding energy and $l=0,1,\cdots$ denotes the angular momentum
quantum number. Of course, on the asymptotically physical grounds, the
wavefunction $\Psi(p)$ satisfies the boundary conditions [3]\\
\begin{center}
$\Psi(p)\approx \left\{ \begin{array}{l}
p^{l+1}~~ for~~ |p|<<1, \\
exp(-\nu p^{3/2})~,~~\nu=\sqrt{\frac{8c}{9m}}(\hbar \omega)^{-1},~~
for~~ |p|>>1.
\end{array} \right.$
\end{center}

A rescal
of the variable $p$ through $p=\sqrt{m \hbar \omega} q$ would, in turn,
lead to a transparent form of Schr\"odinger equation [3]. Strictly,\\
\begin{equation}
\left[-\frac{1}{2}\frac{d^{2}}{dq^{2}}+\frac{l(l+1)}{2q^{2}}+V(q)\right]
\Psi_{n_r,l}(q)=\varepsilon_{n_r,l}\Psi_{n_r,l}(q),
\end{equation}\\
where, $\varepsilon_{n_r,l}=2\acute{E}_{n_r,l}/(\hbar\omega)$, \\
\begin{equation}
V(q)=\frac{1}{\alpha^2}(V_{SRAO}(q)-1),
\end{equation}\\
\begin{equation}
V_{SRAO}(q)=\sqrt{1+\alpha^2q^2},
\end{equation}\\
$\alpha^2=\hbar\omega/(mc^2)$,
and the square - root anharmonic oscillator potential (7) simulates a
quasi - relativistic squeezing of the harmonic oscillator spectrum.

The transition from the parabolic HO well, in standard coordinate
representation, to the hyperbolic shape $V_{SRAO}(q)$, in momentum
representation, is claimed to prove phenomenologically useful and
methodically challenging. Znojil [3] has, therefore, invested several
eligible ( namely, perturbative, variational, Hill - determinant, and
Riccati - Pad\'{e}) methods to construct its bound - states. With the
permission of an anonymous referee of his paper, Znojil has also
reported ( in table 2 of [3]) the referee's results from direct numerical
integrations.

Whilst using a quasi - perturbation prescription ( Eq.(11)in [3]), a loss
of precision occurred at "large" $\alpha=1/2$ ( table 1(b) in [3]).
Upon which
the anonymous referee remarked that it may also cause a loss of the upper -
bound character of the quasi - perturbation prescription. In accordance
with a second referee's remark, being curable by a Pad\'{e} - type
resummation, the loss of boundedness phenomenon may emerge at any $\alpha$. 

To the best of our knowledge, the paper of Znojil [3] is the only available
one in the literature, for the quasi - relativistic harmonic oscillator,
and merits  therefore further considerations.

In this paper we formulate a new method to solve the Fourier transformed
Schr\"odinger equation (4), with $V(q)$ represented by (6).
Our method consists of using
$1/\bar{l}$ as an expansion parameter, where $\bar{l}=l-\beta$,
$l$ is a quantum number, and $\beta$ is a suitable shift introduced, mainly,
to avoid the trivial case $l=0$. The spiritual soundness of the textbook
perturbation theory is therefore engaged. Hence the method should be called
pseudoperturbative shifted - $l$ expansion technique (PSLET).

With the noninteger ( irrational) orbital angular momentum $\bar{l}$,
equation (5) reads\\
\begin{equation}
\left\{-\frac{1}{2}\frac{d^{2}}{dq^{2}}+\tilde{V}(q)\right\}
\Psi_{n_r,l} (q)=\varepsilon_{n_r,l} \Psi_{n_r,l}(q),
\end{equation}\\
\begin{equation}
\tilde{V}(q)=\frac{\bar{l}^{2}+(2\beta+1)\bar{l}
+\beta(\beta+1)}{2q^{2}}+\frac{\bar{l}^2}{Q}V(q).
\end{equation}\\
Where Q is a constant that scales the potential 
$V(q)$ at large - $l$ limit and is set, for any specific choice of $l$
and $n_r$, equal to $\bar{l}^2$ at the end of the calculations [4-9].
And, $\beta$ is to be determined in the sequel.

Our systematic procedure begins with shifting the origin of the
coordinate through\\
\begin{equation}
x=\bar{l}^{1/2}(q-q_{o})/q_{o},
\end{equation}\\
where $q_{o}$ is currently an arbitrary point to perform Taylor expansions
about, with its particular value to be determined.
Expansions about this point yield\\
\begin{equation}
\frac{1}{q^{2}}=\sum^{\infty}_{n=0} (-1)^{n} \frac{(n+1)}{q_{o}^{2}}
 x^{n}\bar{l}^{-n/2},
\end{equation}\\
\begin{equation}
V(x(q))=\sum^{\infty}_{n=0}\left(\frac{d^{n}V(q_{o})}{dq_{o}^{n}}\right)
\frac{(q_{o}x)^{n}}{n!}\bar{l}^{-n/2}.
\end{equation}\\
It should be mentioned here
that the scaled coordinate, equation (10), has no effect on the energy
eigenvalues, which are coordinate - independent. It just facilitates
the calculations of both the energy eigenvalues and eigenfunctions.
It is also convenient to expand $\varepsilon_{n_r,l}$ as\\
\begin{equation}
\varepsilon_{n_r,l} =\sum^{\infty}_{n=-2}
\varepsilon_{n_r,l}^{(n)}\bar{l}^{-n}.
\end{equation}\\
Equation (8) thus becomes\\
\begin{equation}
\left[-\frac{1}{2}\frac{d^{2}}{dx^{2}}+\frac{q_{o}^{2}}{\bar{l}}
\tilde{V}(x(q))\right]
\Psi_{n_r,l}(x)=\frac{q_{o}^2}{\bar{l}}\varepsilon_{n_r,l} \Psi_{n_r,l}(x),
\end{equation}\\
with\\
\begin{eqnarray}
\frac{q_o^2}{\bar{l}}\tilde{V}(x(q))&=&q_o^2\bar{l}
\left[\frac{1}{2q_o^2}+\frac{V(q_o)}{Q}\right]
+\bar{l}^{1/2}\left[-x+\frac{V^{'}(q_o)q_o^3 x}{Q}\right]\nonumber\\
&+&\left[\frac{3}{2}x^2+\frac{V^{''}(q_o) q_o^4 x^2}{2Q}\right]
+(2\beta+1)\sum^{\infty}_{n=1}(-1)^n \frac{(n+1)}{2}x^n \bar{l}^{-n/2}
\nonumber\\
&+&q_o^2\sum^{\infty}_{n=3}\left[(-1)^n \frac{(n+1)}{2q_o^2}x^n
+\left(\frac{d^n V(q_o)}{dq_o^n}\right)\frac{(q_o x)^n}{n! Q}\right]
\bar{l}^{-(n-2)/2}\nonumber\\
&+&\beta(\beta+1)\sum^{\infty}_{n=0}(-1)^n\frac{(n+1)}{2}x^n
\bar{l}^{-(n+2)/2}+\frac{(2\beta+1)}{2},
\end{eqnarray}\\
where the prime of $V(q_o)$ denotes derivative with respect to $q_o$.
Equation (14) is exactly of the type of Schr\"odinger equation 
for one - dimensional anharmonic oscillator\\
\begin{equation}
\left[-\frac{1}{2}\frac{d^2}{dx^2}+\frac{1}{2}\Omega^2 x^2 +\Lambda_o
+P(x)\right]X_{n_r}(x)=\lambda_{n_r}X_{n_r}(x),
\end{equation}\\
where $P(x)$ is a perturbation - like term and $\Lambda_o$ is a 
constant. A simple comparison between Eqs.(14), (15) and (16) implies\\
\begin{equation}
\Lambda_o =\bar{l}\left[\frac{1}{2}+\frac{q_o^2 V(q_o)}{Q}\right]
+\frac{2\beta+1}{2}+\frac{\beta(\beta+1)}{2\bar{l}},
\end{equation}\\
\begin{eqnarray}
\lambda_{n_{r}}&=&\bar{l}\left[\frac{1}{2}+\frac{q_o^2 V(q_o)}{Q}\right]
+\left[\frac{2\beta+1}{2}+(n_r+\frac{1}{2})\Omega\right]\nonumber\\
&+&\frac{1}{\bar{l}}\left[\frac{\beta(\beta+1)}{2}
+\lambda_{n_{r}}^{(0)}\right]
+\sum^{\infty}_{n=2}\lambda_{n_{r}}^{(n-1)}\bar{l}^{-n},
\end{eqnarray}\\
and\\
\begin{equation}
\lambda_{n_{r}} = q_o^2 \sum^{\infty}_{n=-2} \varepsilon_{n_r,l}^{(n)}
\bar{l}^{-(n+1)},
\end{equation}\\
Equations (18) and (19) yield\\
\begin{equation}
\varepsilon_{n_r,l}^{(-2)}=\frac{1}{2q_o^2}+\frac{V(q_o)}{Q}
\end{equation}\\
\begin{equation}
\varepsilon_{n_r,l}^{(-1)}=\frac{1}{q_o^2}\left[\frac{2\beta+1}{2}
+(n_r +\frac{1}{2})\Omega\right]
\end{equation}\\
\begin{equation}
\varepsilon_{n_r,l}^{(0)}=\frac{1}{q_o^2}\left[ \frac{\beta(\beta+1)}{2}
+\lambda_{n_r}^{(0)}\right]
\end{equation}\\
\begin{equation}
\varepsilon_{n_r,l}^{(n)}=\lambda_{n_r}^{(n)}/q_o^2  ~~;~~~~n \geq 1.
\end{equation}\\
Here $q_o$ is chosen to minimize $\varepsilon_{n_r,l}^{(-2)}$, i. e.\\
\begin{equation}
\frac{d\varepsilon_{n_r,l}^{(-2)}}{dq_o}=0~~~~
and~~~~\frac{d^2 \varepsilon_{n_r,l}^{(-2)}}{dq_o^2}>0,
\end{equation}\\
which in turn gives, with $\bar{l}=\sqrt{Q}$,\\
\begin{equation}
l-\beta=\sqrt{q_{o}^{3}V^{'}(q_{o})}.
\end{equation}\\
Consequently, the second term in Eq.(15) vanishes and the first term adds 
a constant to the energy eigenvalues.

The next leading correction to the energy series,
 $\bar{l}\varepsilon_{n_r,l}^{(-1)}$,
consists of a constant term and the exact eigenvalues of the unperturbed
harmonic oscillator potential $\Omega^2x^2/2$. 
The shifting parameter $\beta$ is determined by choosing
$\bar{l}\varepsilon_{n_r,l}^{(-1)}$=0. Hence\\
\begin{equation}
\beta=-\left[\frac{1}{2}+(n_{r}+\frac{1}{2})\Omega\right],
\end{equation}\\
where\\
\begin{equation}
\Omega=\sqrt{3+\frac{q_o V^{''}(q_o)}{V^{'}(q_o)}}.
\end{equation}\\

Then equation (15) reduces to\\
\begin{equation}
\frac{q_o^2}{\bar{l}}\tilde{V}(x(q))=
q_o^2\bar{l}\left[\frac{1}{2q_o^2}+\frac{V(q_o)}{Q}\right]+
\sum^{\infty}_{n=0} v^{(n)}(x) \bar{l}^{-n/2},
\end{equation}\\
where\\
\begin{equation}
v^{(0)}(x)=\frac{1}{2}\Omega^2 x^2 + \frac{2\beta+1}{2},
\end{equation}\\
\begin{equation}
v^{(1)}(x)=-(2\beta+1) x - 2x^3 + \frac{q_o^5 V^{'''}(q_o)}{6 Q} x^3,
\end{equation}\\
and for $n \geq 2$\\
\begin{eqnarray}
v^{(n)}(x)&=&(-1)^n (2\beta+1) \frac{(n+1)}{2} x^n
+ (-1)^{n} \frac{\beta(\beta+1)}{2} (n-1) x^{(n-2)}\nonumber\\
&+& \left[(-1)^{n} \frac{(n+3)}{2}
+ \frac{q_o^{(n+4)}}{Q(n+2)!} \frac{d^{n+2} V(q_o)}{dq_o^{n+2}}\right]
x^{n+2}.
\end{eqnarray}\\
Equation (14) thus becomes\\
\begin{eqnarray}
&&\left[-\frac{1}{2}\frac{d^2}{dx^2} + \sum^{\infty}_{n=0} v^{(n)}
\bar{l}^{-n/2}\right]\Psi_{n_r,l} (x)= \nonumber\\
&& \left[\frac{1}{\bar{l}}\left(\frac{\beta(\beta+1)}{2}
+\lambda_{n_r}^{(0)}\right) 
+ \sum^{\infty}_{n=2} \lambda_{n_r}^{(n-1)}
\bar{l}^{-n} \right] \Psi_{n_r,l}(x).
\end{eqnarray}\\

When setting the nodeless, $n_r = 0$, wave functions as \\
\begin{equation}
\Psi_{0,l}(x(q)) = exp(U_{0,l}(x)),
\end{equation}\\
equation (32) is readily transformed into the following Riccati equation:\\
\begin{eqnarray}
-\frac{1}{2}[ U^{''}(x)+U^{'}(x)U^{'}(x)]
+\sum^{\infty}_{n=0} v^{(n)}(x) \bar{l}^{-n/2}
&=&\frac{1}{\bar{l}} \left( \frac{\beta(\beta+1)}{2}
+ \lambda_{0}^{(0)}\right)\nonumber\\
&&+\sum^{\infty}_{n=2} \lambda_{0}^{(n-1)} \bar{l}^{-n}.
\end{eqnarray}\\
Hereinafter, we shall use $U(x)$ instead of $U_{0,l}(x)$ for simplicity,
and the prime of $U(x)$ denotes derivative with respect to $x$. It is
evident that this equation admits solution of the form \\
\begin{equation}
U^{'}(x)=\sum^{\infty}_{n=0} U^{(n)}(x) \bar{l}^{-n/2}
+\sum^{\infty}_{n=0} G^{(n)}(x) \bar{l}^{-(n+1)/2},
\end{equation}\\
where\\
\begin{equation}
U^{(n)}(x)=\sum^{n+1}_{m=0} D_{m,n} x^{2m-1} ~~~~;~~~D_{0,n}=0,
\end{equation}\\
\begin{equation}
G^{(n)}(x)=\sum^{n+1}_{m=0} C_{m,n} x^{2m}.
\end{equation}\\
Substituting equations (35) - (37) into equation (34) implies\\
\begin{eqnarray}
&-&\frac{1}{2} \sum^{\infty}_{n=0}\left[U^{(n)^{'}} \bar{l}^{-n/2}
+ G^{(n)^{'}} \bar{l}^{-(n+1)/2}\right] \nonumber\\
&-&\frac{1}{2} \sum^{\infty}_{n=0} \sum^{\infty}_{p=0}
\left[ U^{(n)}U^{(p)} \bar{l}^{-(n+p)/2}
+G^{(n)}G^{(p)} \bar{l}^{-(n+p+2)/2}
+2 U^{(n)}G^{(p)} \bar{l}^{-(n+p+1)/2}\right]\nonumber\\
&+&\sum^{\infty}_{n=0}v^{(n)} \bar{l}^{-n/2}
=\frac{1}{\bar{l}}\left(\frac{\beta(\beta+1)}{2}+\lambda_{0}^{(0)}\right)
+\sum^{\infty}_{n=2} \lambda_{0}^{(n-1)} \bar{l}^{-n},
\end{eqnarray}\\
where primes of $U^{(n)}(x)$ and $G^{(n)}(x)$ denote derivatives
with respect to $x$. Equating the coefficients of the same powers of
$\bar{l}$ and $x$, respectively, ( of course the other way around would 
work equally well) one obtains\\
\begin{equation}
-\frac{1}{2}U^{(0)^{'}} - \frac{1}{2}  U^{(0)} U^{(0)} + v^{(0)} = 0,
\end{equation}\\
\begin{equation}
U^{(0)^{'}}(x) = D_{1,0} ~~~;~~~~D_{1,0}=-\Omega,
\end{equation}\\
and integration over $dx$ yields\\
\begin{equation}
U^{(0)}(x)=-\Omega x.
\end{equation}\\
Similarly,\\
\begin{equation}
-\frac{1}{2}[U^{(1)^{'}} + G^{(0)^{'}}] - U^{(0)}U^{(1)} - U^{(0)}G^{(0)}
+v^{(1)}=0,
\end{equation}\\
\begin{equation}
U^{(1)}(x)=0,
\end{equation}\\
\begin{equation}
G^{(0)}(x)=C_{0,0}+C_{1,0}x^2,
\end{equation}\\
\begin{equation}
C_{1,0}=-\frac{B_{1}}{\Omega},
\end{equation}\\
\begin{equation}
C_{0,0}=\frac{1}{\Omega}(C_{1,0}+2\beta+1),
\end{equation}\\
\begin{equation}
B_{1}=-2+\frac{q_o^5}{6Q}\frac{d^3 V(q_o)}{dq_o^3},
\end{equation}\\
\begin{eqnarray}
&&-\frac{1}{2}[U^{(2)^{'}} + G^{(1)^{'}}]
- \frac{1}{2}\sum^{2}_{n=0}U^{(n)}U^{(2-n)}-\frac{1}{2}G^{(0)}G^{(0)}
\nonumber\\
&&-\sum^{1}_{n=0}U^{(n)}G^{(1-n)} + v^{(2)}
= \frac{\beta(\beta+1)}{2} + \lambda_{0}^{(0)},
\end{eqnarray}\\
\begin{equation}
U^{(2)}(x)=D_{1,2}x + D_{2,2}x^3,
\end{equation}\\
\begin{equation}
G^{(1)}(x)=0,
\end{equation}\\
\begin{equation}
D_{2,2}=\frac{1}{\Omega}(\frac{C_{1,0}^2}{2}-B_{2})
\end{equation}\\
\begin{equation}
D_{1,2}=\frac{1}{\Omega}(\frac{3}{2}D_{2,2}+C_{0,0}C_{1,0}
-\frac{3}{2}(2\beta+1)),
\end{equation}\\
\begin{equation}
B_{2}=\frac{5}{2}+\frac{q_o^6}{24Q}\frac{d^4V(q_o)}{dq_o^4},
\end{equation}\\
\begin{equation}
\lambda_{0}^{(0)} = -\frac{1}{2}(D_{1,2}+C_{0,0}^2).
\end{equation}\\
$\cdots$ and so on. Thus, one can calculate the energy 
eigenvalue and the eigenfunctions from the knowledge of $C_{m,n}$
and $D_{m,n}$ in a hierarchical manner.
Nevertheless, the procedure just described is suitable
 for systematic calculations
using software packages (such as MATHEMATICA, MAPLE, or REDUCE) to determine
the energy eigenvalue and eigenfunction corrections up to any order of the
pseudoperturbation series. 

It should be mentioned that the energy series, Eq.(13), could appear
convergent, divergent, or asymptotic. However, one can still
calculate the eigenenergies to a very good accuracy by forming the 
sophisticated Pad\'{e} approximants to the energy series [11]. 
The energy series, Eq.(13), is calculated up to 
$\varepsilon_{0,l}^{(8)}/\bar{l}^{8}$ by
\begin{equation}
\varepsilon_{0,l}=\bar{l}^{2}\varepsilon_{0,l}^{(-2)}
+\varepsilon_{0,l}^{(0)}+\cdots
+\varepsilon_{0,l}^{(8)}/\bar{l}^{8}+O(1/\bar{l}^{9}),
\end{equation}\\
and with the $P_{N}^{N}(1/\bar{l})$ and $P_{N}^{N+1}(1/\bar{l})$
Pad\'{e} approximants it becomes\\
\begin{equation}
\varepsilon_{0,l}[N,N]=\bar{l}^{2}\varepsilon_{0,l}^{(-2)}
+P_{N}^{N}(1/\bar{l}).
\end{equation}\\
and\\
\begin{equation}
\varepsilon_{0,l}[N,N+1]=\bar{l}^{2}\varepsilon_{0,l}^{(-2)}
+P_{N}^{N+1}(1/\bar{l}).
\end{equation}\\
Our strategy and prescription are therefore clear.

Let us now consider the Fourier transformed equation, Eq.(4), with the
rescaled variable $p = \sqrt{m\hbar\omega}q$, represented by equations
(5)-(7). The substitution of Eq.(6) in (26), for $n_r=0$, implies\\
\begin{equation}
\beta=-\frac{1}{2}(1+\Omega)~~;~~~~
\Omega=\sqrt{\frac{4+3\alpha^2 q_o^2}{1+\alpha^2 q_o^2}}
\end{equation}\\
Eq.(25) thus reads\\
\begin{equation}
l+\frac{1}{2}(1+\Omega)
=q_o^2\sqrt{\frac{1}{\sqrt{1+\alpha^2 q_o^2}}}.
\end{equation}\\
Equation (59) is explicit in $q_o$ and evidently a closed form solution
for $q_o$ is hard to find, though almost impossible. However, numerical
solutions are feasible. Once $q_o$ is determined the coefficients
$C_{m,n}$ and $D_{m,n}$ are obtained in a sequential manner. Consequently,
the eigenvalues, Eq.(55), and eigenfunctions, Eqs.(35)-(37), are calculated
in the same batch for each value of $\alpha$, and $l$.

In order to make remediable analysis of our results we have calculated
the first ten terms of the energy series. The effect of each term has
been taken into account. We have also computed the Pad\'{e} approximants
$\varepsilon_{0,l}[N,M]$ for $N=2,3,4$ and $M=2,3,4,5$. Therefore,
the stability of the energy series and that of the sequence of Pad\'{e}
approximants are in point.

Table 1 shows PSLET results for the ground - state energies
$\varepsilon_{0,0}$, covering a wide range of the anharmonicity $\alpha$,
along with the exact ones from direct numerical integration method ( DNI),
carried out by the anonymous referee of paper [3]. To avoid exhaustive
numbers of tables we do not list Znojil's results. However, we will just
refer to them. A comparison between PSLET and DNI results implies
excellent agreements. The nice trend of stability of the energy series (55)
(i.e., a signal of nice course of convergence.) is well pronounced. The
effect of the higher - order corrections on the first few terms of the
energy series bears this out.

In contrast with Znojil's results ( table 1(b) in [3]) for "large"
$\alpha$=1/3 and $\alpha$=1/2, via a quasi - perturbative prescription
( Eq.(11) in [3]), there is no indication that our series will blow up
at higher - orders and ( since, mainly,) our expansion parameter
$(1/\bar{l})$ is less than one for all values of $\alpha$
reported in the text. Of course there is always the contribution of the
$(K+1)$ - term, but so far our prescription is performing so good.
Whilst Znojil's prescription marks nice stability
for small $\alpha$, sever oscillations of his series occur at low -
order, especially for "large" $\alpha$=1/2, causing in effect a breakdown
in the boundedness character of his prescription. Although this phenomenon
is curable by resummation tools like the sophisticated Pad\'{e} approximants,
as suggested by the second referee of his paper [3], however, this, in our
opinion, shall not dramatically cure the loss of precision in Znojil's
results ( table 1(b) in [3]), to be documented in the sequel.

Switching to alternative methods for independent checks of his numerical
predictions, Znojil used the Hill determinant and Riccati - Pad\'{e}
methods. In the light of his experience in the Hill determinant, an
onset of convergence is clearly manifested ( table 3 in [3]), but larger
dimensions and/or improved elementary convergence factor would be
necessary to reach the domain of more satisfactory numerical precision.
Upon his experience, moreover, in a slightly more complicated ( compared
to the Hill determinant) Riccati - Pad\'{e} method ( RPM), the
11- dimensional T\"oplitz determinants have offered very satisfactory
precision ( table 4 in [3]). However, a typical bizarre characteristic
of the RPM is well documented [3,11]. Namely, it leads to a number of
clustered  solutions, for a given value of the coupling $\alpha$, resulting
from the existence of several eligible physical roots of the Hankel [11]
or T\"oplitz [3] determinants. Yet the ambiguity of these
roots increases with the dimensional growth of the determinants. Although
clustering is a good signal of being close to a physical root, a decision on
which of these roots is the best has to be made. So far, to the best of
our knowledge, a general way of establishing this property has not been
found.

The effect of the angular momentum quantum number $l$ on the stability,
hence on convergence, and precision is reported in table 2 for $\alpha$=1/2.
Confidently, one concludes that better convergence and more precise numerical
results are obtained as $l$ grows up. Similar effect should be expected
from the nodal quantum number $n_r$; $l$ and $n_r$ have almost identical
effects on our pseudoperturbative expansion parameter $\bar{l}$.

The stability of the sequence of Pad\'{e} approximants ( table 3) is
fascinating. Although there is no signal that our series will blow up,
table 1 marks this fact, the effect of Pad\'{e} approximants on
precision is limited. For a fixed $\alpha$, say 1/2, more precision
is obtained via Pad\'{e} approximants as $l$ increases ( table 4).
Adhered to the conventional practice of perturbative calculations ( i.e.
only a few terms of a " most useful" perturbation series reveal the
important features of the solution before a state of exhaustion is
reached.), we list PSLET results ( table 5) from the first 6 terms of our
energy series with $\varepsilon_{0,0}[3,3]$ Pad\'{e} approximant.
Compared to those from DNI, our results are readily satisfactory.

To summarize, we have used a new methodical proposal to investigate
the bound - states of the quasi - relativistic harmonic oscillator. Using
the perturbation expansion parameter $1/\bar{l}$, we have demonstrated
that our apparently artificial perturbation recipe PSLET is convincingly
powerful and methodically practical.

Perhaps it should be noted that for each entry in tables 1-5 one can
construct the wavefunction from the knowledge of $C_{m,n}$ and $D_{m,n}$.
However, such a study lies beyond the scope of our methodical proposal.

In addition to Znojil's interpretation of the QHO Hamiltonian (3) that
it leads, in effect, to the formally correct relativistic Dirac equation,
we have shown that it could as well represent a Klein - Gordon particle
in a parabolic well. Precisely, the 4 - vector potential
$eA_0=-Ze^2(3-r^2/a)/2a$, or in short $eA_0=A+Br^2$, represents an improved
approximation for a realistic pionic atom, hence the Hamiltonian in Eq.(3)
addresses the Klein - Gordon Hamiltonian  for the potential of a
homogeneously charged sphere [13].

The applicability of our recipe extends beyond the present quasi
- relativistic harmonic oscillator model. Some applications are in
order. The eigenstates of a hydrogenic impurity in a spherical quantum
dot (QD) [14]. Quasi - two - dimensional QD helium [15]. Two - electrons
QD in a magnetic field [16]. Excitons in harmonic QD [17]. Hydrogenic
impurity or heavy excitons in arbitrary magnetic field [18,19], $\cdots$ etc.

\newpage
\begin{table}
\begin{center}
\caption{ Ground - state energies
, where $K$ represents the first $K$ - terms
of Eq.(54) and DNI from direct numerical integration [3].}
\vspace{0.5cm}
\begin{tabular}{|c|c|c|c|c|}
\hline \hline
$K$ & $\alpha =1/100$ & $\alpha=1/20$ & $\alpha=1/10$ & $\alpha=1/5$
\\ \hline
1 & 2.999906 257850 8 & 2.997661 14471 & 2.990702 721 & 2.963706 98 \\
2 & 2.999906 259959 4 & 2.997662 45159 & 2.990723 092 & 2.964001 14 \\
3 & 2.999906 259959 1 & 2.997662 44641 & 2.990722 775 & 2.963984 17 \\
4 & 2.999906 259959 1 & 2.997662 44644 & 2.990722 783 & 2.963985 58 \\
5 & 2.999906 259959 1 & 2.997662 44644 & 2.990722 782 & 2.963985 42 \\
6 & 2.999906 259959 1 & 2.997662 44644 & 2.990722 782 & 2.963985 45 \\
7 & 2.999906 259959 1 & 2.997662 44631 & 2.990722 775 & 2.963985 06 \\
8 & 2.999906 259959 1 & 2.997662 44635 & 2.990722 777 & 2.963985 17 \\
9 & 2.999906 259959 1 & 2.997662 44635 & 2.990722 777 & 2.963985 17 \\
10& 2.999906 259959 1 & 2.997662 44634 & 2.990722 776 & 2.963985 14\\
 \hline
DNI   &  $----$         & 2.997662 44644 & 2.990722 782 & 2.963985 44 \\
\hline \hline
& $\alpha=1/4$ & $\alpha=1/3$ & $\alpha=1/2$ & $\alpha=2$ \\ \hline
1 & 2.944289 62 & 2.904543 & 2.80482 & 1.9189 \\ 
2 & 2.944955 82 & 2.906342 & 2.81083 & 1.9331 \\ 
3 & 2.944899 04 & 2.906100 & 2.80951 & 1.9334 \\ 
4 & 2.944905 92 & 2.906145 & 2.80987 & 1.9323 \\ 
5 & 2.944904 81 & 2.906135 & 2.80977 & 1.9319 \\ 
6 & 2.944905 03 & 2.906138 & 2.80979 & 1.9322 \\ 
7 & 2.944903 79 & 2.906132 & 2.80978 & 1.9328 \\ 
8 & 2.944904 11 & 2.906134 & 2.80977 & 1.9329 \\ 
9 & 2.944904 11 & 2.906134 & 2.80977 & 1.9329 \\ 
10& 2.944904 03 & 2.906133 & 2.80975 & 1.9315 \\
\hline
DNI   & 2.944904 99 & 2.906136 & 2.809786 & 1.932334 \\
\hline \hline
\end{tabular}
\end{center}
\end{table}
\newpage
\begin{table}
\begin{center}
\caption{ The effect of the angular momentum quantum number $l$
on convergence and precision for $\alpha=1/2$.}
\vspace{0.5cm}
\begin{tabular}{|c|c|c|c|c|}
\hline \hline
$K$ & $l=1$ & $l=5$ & $l=10$ & $l=20$ \\ \hline
1 & 4.569573 & 10.977406 1 & 17.963301 722 & 29.948955 412735 \\
2 & 4.575998 & 10.981520 1 & 17.965505 044 & 29.949880 189612 \\
3 & 4.575031 & 10.981360 3 & 17.965487 822 & 29.949881 669238 \\ 
4 & 4.575173 & 10.981349 8 & 17.965484 484 & 29.949881 347421 \\ 
5 & 4.575167 & 10.981352 9 & 17.965484 566 & 29.949881 340115 \\ 
6 & 4.575155 & 10.981352 8 & 17.965484 596 & 29.949881 340512 \\ 
7 & 4.575164 & 10.981352 7 & 17.965484 595 & 29.949881 340559 \\ 
8 & 4.575161 & 10.981352 7 & 17.965484 595 & 29.949881 340559 \\ 
9 & 4.575161 & 10.981352 7 & 17.965484 595 & 29.949881 340559 \\ 
10& 4.575162 & 10.981352 7 & 17.965484 595 & 29.949881 340559 \\
\hline \hline
\end{tabular}
\end{center}
\end{table}
\newpage
\begin{table}
\begin{center}
\caption{ The effect of Pad\'{e} approximants on convergence and precision.}
\vspace{0.5cm}
\begin{tabular}{|c|c|c|c|c|}
\hline \hline
$\varepsilon_{0,0} [N,M]$
& $\alpha=1/100$ & $\alpha=1/10$ & $\alpha=1/3$ & $\alpha=2$ \\
\hline
$\varepsilon_{0,0} [2,2]$
& 2.999906 259959 1 & 2.990722 78 & 2.90614 & 1.9319 \\
$\varepsilon_{0,0}[2,3]$
& 2.999906 259959 1 & 2.990722 78 & 2.90614 & 1.9323 \\
$\varepsilon_{0,0}[3,3]$
& 2.999906 259959 1 & 2.990722 78 & 2.90614 & 1.9334 \\
$\varepsilon_{0,0}[3,4]$
& 2.999906 259959 1 & 2.990722 78 & 2.90614 & 1.9325 \\
$\varepsilon_{0,0}[4,4]$
& 2.999906 259959 1 & 2.990722 78 & 2.90613 & 1.9326 \\
$\varepsilon_{0,0}[4,5]$
& 2.999906 259959 1 & 2.990722 78 & 2.90613 & 1.9326 \\
\hline \hline
\end{tabular}
\end{center}
\end{table}
\newpage
\begin{table}
\begin{center}
\caption{ The effect of $l$ and Pad\'{e} approximants on convergence
and precision for $\alpha=1/2$.}
\vspace{0.5cm}
\begin{tabular}{|c|c|c|c|}
\hline \hline
$\varepsilon_{0,l}[N,M]$ & $l=0$ & $l=1$ & $l=3$ \\ \hline
$\varepsilon_{0,l}[2,2]$ & 2.809788 & 4.575157 & 7.893680  \\ 
$\varepsilon_{0,l}[2,3]$ & 2.809826 & 4.575161 & 7.893681 \\ 
$\varepsilon_{0,l}[3,3]$ & 2.809783 & 4.575161 & 7.893681 \\ 
$\varepsilon_{0,l}[3,4]$ & 2.809767 & 4.575161 & 7.893681 \\ 
$\varepsilon_{0,l}[4,4]$ & 2.809790 & 4.575161 & 7.893681 \\ 
$\varepsilon_{0,l}[4,5]$ & 2.809797 & 4.575161 & 7.893681 \\ 
\hline \hline
        & $l=5$ & $l=10$ & $l=20$ \\ \hline
$\varepsilon_{0,l}[2,2]$
& 10.981352 490 & 17.965484 570 & 29.949881 340088 \\
$\varepsilon_{0,l}[2,3]$
& 10.981352 767 & 17.965484 596 & 29.949881 340530 \\
$\varepsilon_{0,l}[3,3]$
& 10.981352 715 & 17.965484 595 & 29.949881 340562 \\
$\varepsilon_{0,l}[3,4]$
& 10.981352 712 & 17.965484 587 & 29.949881 340559 \\
$\varepsilon_{0,l}[4,4]$
& 10.981352 712 & 17.965484 595 & 29.949881 340559 \\
$\varepsilon_{0,l}[4,5]$
& 10.981352 712 & 17.965484 595 & 29.949881 340559 \\
\hline \hline
\end{tabular}
\end{center}
\end{table}
\newpage
\begin{table}
\begin{center}
\caption{ Comparison between PSLET, collecting the first 6 terms of the
energy series, $\varepsilon_{0,0}[3,3]$ Pad\'{e} approximant
and the results from direct numerical integration (DNI) [3].}
\vspace{0.5cm}
\begin{tabular}{|c|c|c|c|}
\hline \hline
$\alpha$ & $K=6$ & $\varepsilon_{0,0}[3,3]$ & DNI [3] \\ \hline
1/100 & 2.999906 259959 1 & 2.999906 2599591 & - \\ 
1/20  & 2.997662 44644    & 2.997662 44644 & 2.997662 44644 \\ 
1/10  & 2.990722 78231    & 2.990722 78231 & 2.990722 78232 \\ 
1/5   & 2.963985 445    & 2.963985 441 & 2.963985 44193 \\ 
1/4   & 2.944905 033    & 2.944904 983 & 2.944904 99229 \\ 
1/3   & 2.906137 610    & 2.906136 824 & 2.906136 36892 \\ 
1/2   & 2.809786 91    & 2.809783 442 & 2.809786 32134 \\ 
2     & 1.932185     & 1.933444 & 1.932334 34201 \\
\hline \hline
\end{tabular}
\end{center}
\end{table}
\end{document}